\documentclass{article}
\usepackage{hiph-art}
\volnumber{19} \issuenumber{1} \edyear{2004}                             %|
\frompage{000} \topage{000}                                              %|
\recrevdate{20 July 2004}                                              %|
\def\rightmark{Energy dependence of transverse flow in heavy ion
collisions}
\def\leftmark{P. Csizmadia and P. L{\'e}vai}
 
\title{Energy dependence of transverse quark flow\\
in heavy ion collisions}
\authors{ 
{P{\'e}ter Csizmadia$^1$ and P{\'e}ter L{\'e}vai$^1$ %
\index{Csizmadia, P.} % Abbreviated names of the author(s),
\index{L{\'e}vai, P.} % to be inserted for use in the volume index
}\\[2.812mm]
{\normalsize
\hspace*{-8pt}$^1$ KFKI Research Institute for Particle and Nuclear Physics,\\
PO. Box 49, Budapest 1525, Hungary\\[0.2ex] 
}}
 
\abstract{Energy dependence of  quark transverse flow carries 
  information about dynamical properties (equation of state, initial
  conditions) of deconfined matter 
  produced in heavy ion collisions. We assume quark-antiquark matter 
  formation in Pb+Pb collisions at CERN SPS and Au+Au collisions at RHIC
  energies and determine quark transverse flow at the critical
  temperature of the quark-hadron phase transition. 
  Coalescence of massive quarks is calculated in the MICOR hadronization
  model and hadronic final state effects are considered using the GROMIT
  cascade program. Comparing theoretical results to  data,
  transverse flow values are determined and  energy dependence is discussed.}


 %see: http://www.aip.org/pacs/
 
\makeindex
\begin{document}
 
\maketitle

\section{Introduction}\label{intro}
In heavy ion collisions the formation of quark-gluon plasma (QGP) is
expected at RHIC energies and above, however at lower energies we 
may not see a fully deconfined state.  Although the precursors of the QGP 
state has been seen at CERN SPS \cite{CERN-EvidenceForNewStateOfMatter}, it is
still controversial whether this new state of matter could really appear in
this energy range.  To retrieve information about the energy dependence of
quark matter formation, experimental data must be compared to model
calculations based on deconfinement in a wide energy range. 
Experimental data on transverse momentum spectra for different
particle species existed at
$\sqrt{s_\mathrm{NN}}=17.3\,\mathrm{GeV}$ \cite{SPS158}, $130\,\mathrm{GeV}$
\cite{RHIC130-phenix,RHIC130-star-pi}
and $200\,\mathrm{GeV}$ \cite{RHIC200-phenix,RHIC200-star}. 
Recently, NA49 Collaboration published new data
at energies $\sqrt{s_\mathrm{NN}}=8.8\,\mathrm{GeV}$ and $12.3\,\mathrm{GeV}$
($E_\mathrm{lab}=40\,A\,\mathrm{GeV}$  and $80\,A\,\mathrm{GeV}$) 
\cite{na49-pi+K,na49-p,na49-phi+Omega,na49-Lambda},
which extended the accessible region.
\newpage

In this paper we are interested in the transverse dynamics of heavy ion 
collisions.  Transverse collective flow is directly related to the pressure
of the central region with high energy density. Thus any jump
(or discontinuity) in the flow can indicate modifications in 
the equation of state. We investigate hadronic transverse momentum spectra 
in the energy range $\sqrt{s_\mathrm{NN}}=8.8-200 \,\mathrm{GeV}$
and determine quark transverse flow before hadronization.

We assume the formation of deconfined matter with high energy density 
at all bombarding energies in the above mentioned energy range.  
Introducing critical temperature $T_c$ for the quark-hadron phase transition,
we assume a thermally equilibrated quark matter.
The analysis of lattice-QCD results indicates
the formation of a (constituent) quark-antiquark dominated deconfined matter
at phase transition~\cite{LevHeinz98}.
This quark-antiquark matter can be described by macroscopical parameters as
temperature, particle densities and flow velocities. 
Here we consider $T_c$ to be a uniform phase transition temperature at all
collision energies. Quark densities 
and flow velocities are energy dependent.
Since quark flow cannot be measured directly, we determine it
indirectly from hadronic momentum spectra. We apply hadronization models
based on coalescence of massive quarks and antiquarks with
small relative momentum (ALCOR model~\cite{ALCOR,ALCOR00} to 
reproduce particle yields and MICOR~\cite{MICOR,MICOR02}
to reproduce transverse momentum spectra at low momenta, $p_T < 2$ GeV). 
Our analysis is based on the MICOR model implementing different quark flow 
velocities and calculated results are compared to measured data.
In parallel we will investigate the influence of hadronic rescattering
on hadronic flow by the GROMIT cascade program~\cite{GROMIT}.

Recently parton coalescence models were applied successfully at RHIC energies
in the intermediate momentum region ($2 < p_T < 6$ GeV) to explain 
the measured anomalous $p^+/\pi^+$ and ${\overline p}^-/\pi^-$ ratios
and reproduce the measured transverse momentum 
spectra~\cite{recHwa,recFries,coalGKL}.
In parallel, application of parton coalescence can reproduce 
the measured properties of the elliptic flow at RHIC energies
\cite{coalGKL,coalMolnar,coalKolb}.
The success of these calculations supports the application of
quark coalescence in a wide momentum region. The MICOR model
offers the possibility to investigate coalescence,
especially flow phenomena, in  low-$p_T$ region.

\section{Calculations with the MICOR model}

In our model the expanding quark-antiquark matter cools down to
critical temperature, $T_c=170\,\mathrm{MeV}$, 
where hadronization starts. The longitudinally expanding
deconfined matter is limited to the coordinate rapidity region
$\eta = [-\eta_\mathrm{max}, +\eta_\mathrm{max}]$. Inside this region
we assume constant (anti)quark densities, $dN_{i \bar i}/d\eta$.
We use the following notation for the 
numbers of newly produced quark-antiquark pairs at $T_c$:
%including the strange quark-antiquark pairs with the $f_s$ factor:
\begin{eqnarray}
N_{u\bar u}\,\ =\,\ N_{d\bar d}\,\ =\,\ N_{q\bar q}/2 \ \ , \quad
& & N_{s\bar s}\,\ =\,\ f_s N_{q\bar q} \ \ .
\end{eqnarray}

\begin{figure}[htb]
                 \insertplot{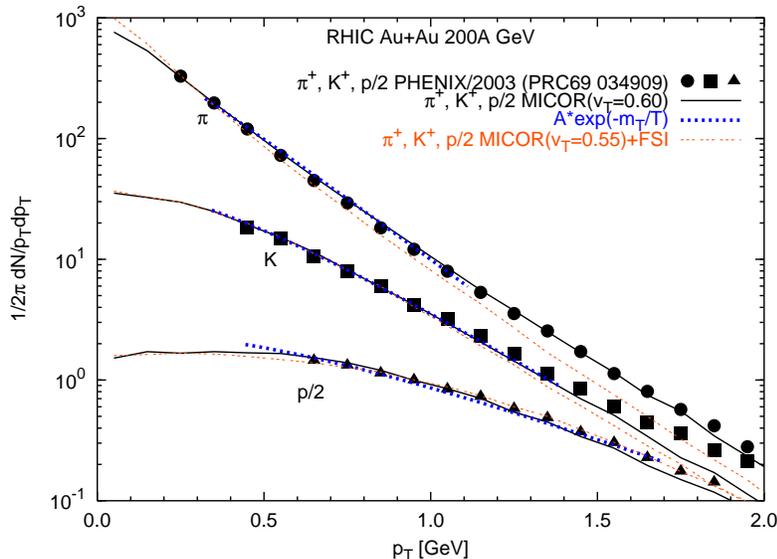}
\vspace*{-8mm}
\caption[]{Transverse momentum spectra for $\pi^+, \ K^+$, and $p$
at $\sqrt{s_{NN}}= 200\,\mathrm{GeV}$ in AuAu collisions
(dots, squares and triangulars) \cite{RHIC200-phenix}.
Full lines indicate  quark coalescence results at $v_T=0.60$,
neglecting final state interactions. Thick dashed lines show
the applied exponential fits to extract slope parameters.
Thin dotted lines indicate the MICOR results at a smaller transverse flow,
$v_T=0.55$, but including final state interactions (FSI).
}
\label{fig-slopes-RHIC}
\end{figure}

In the MICOR model 
final hadron numbers are scaled linearly on the initial quark numbers.
This linear scaling remains valid for the rapidity densities also.
In the ALCOR calculations for
mid-rapidity data one can see the collision energy dependence of the
number of newly produced quark-antiquark numbers~\cite{ALCOR00},
which can be described approximately as 
$dN_{q{\overline q}}/dy = 130 \cdot \ln (\sqrt{s_{NN}}/2.5)$ ~\cite{Lev04}.

The MICOR model covers wide space-time rapidity region and
it reproduces measured mid-rapidity momentum spectra 
by the following parameters:
$\eta_\mathrm{max}=1.9$, $N_{q\bar q}=720$, $f_s=0.35$ at 
$\sqrt{s_{NN}}=17.3$ GeV (SPS), and
$\eta_\mathrm{max}=4$, $N_{q\bar q}=3800$, $f_s=0.26$ at 
$\sqrt{s_{NN}}=200$ GeV (RHIC).
At different SPS energies the above $N_{q \overline q}$ value is scaled by
the measured charged particle yield, as well as at RHIC energies.
(A detailed analysis will be published elsewhere~\cite{CSP04}.)

The inverse slopes for transverse hadronic momentum spectra depend on
the quark flow but not on the quark number.
Secondary interactions among hadrons depend
on absolute hadron numbers.
We can set all above mentioned parameters of MICOR, but left only one
parameter free, namely the quark transverse flow, $v_T$.
We investigate the $v_T$ dependence of hadronic transverse momentum spectra
of different hadronic species.
In this comparison, as one of the possibilities,
we will assume fast hadronization and neglect
final state interactions.

\begin{figure}[htb]
                 \insertplot{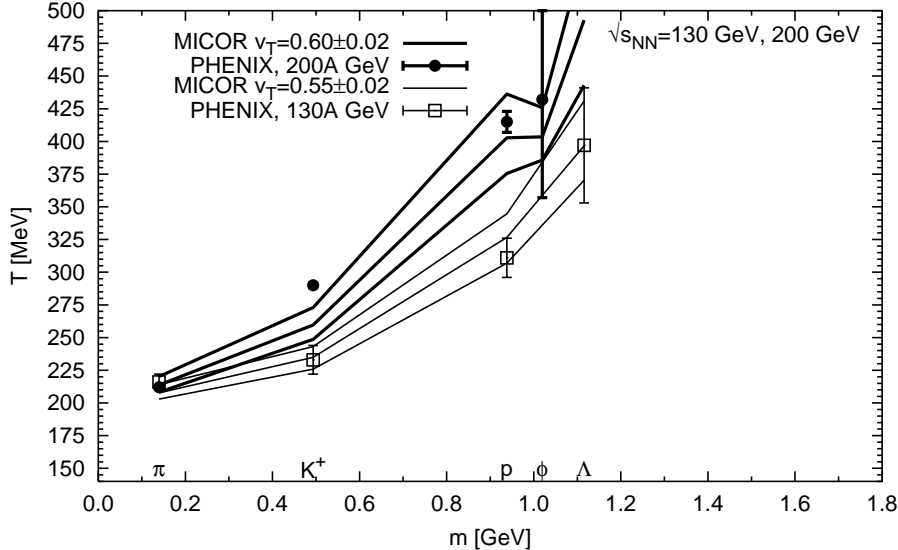}
\vspace*{-8mm}
\caption[]{Inverse slopes from MICOR model and RHIC experiment.
  MICOR points are connected by thin lines for $v_T=0.55\pm0.02$ flow
  velocities and thick lines for $v_T=0.60\pm0.02$.
  $\sqrt{s_{NN}}=130\,\mathrm{GeV}$ data points from
  PHENIX \cite{RHIC130-phenix} are denoted by open squares, 
  $200\,\mathrm{GeV}$ points from PHENIX \cite{RHIC200-phenix} 
  are denoted by dots.}
\label{fig-slopes0-RHIC}
\end{figure}

We consider $\pi^+, \ K^+$, and $p$ 
transverse momentum spectra 
at $\sqrt{s_{NN}}=200$ GeV \cite{RHIC200-phenix}
in AuAu collisions, see Figure \ref{fig-slopes-RHIC}.
We expect an energetic collision with a fast expansion and fast hadronization.
In this case we can compare the MICOR results (full lines)
to the experimental data directly. 
The agreement between MICOR ($v_t=0.6$) and the data points
at $0.2 < p_T < 1.5$ GeV indicates the validity of the above expectation.  
Dashed lines show  exponential fits to the calculated spectra. 

On the other hand,
we are able to switch on hadronic interactions in the final state
by using the GROMIT transport code~\cite{GROMIT}.
However, GROMIT needs space-time information, which was integrated out
in the MICOR model. Here a cylindrical symmetric matter was assumed
with uniform densities, longitudinal Bjorken flow and constant transverse 
flow.

\begin{figure}[ht]
                 \insertplot{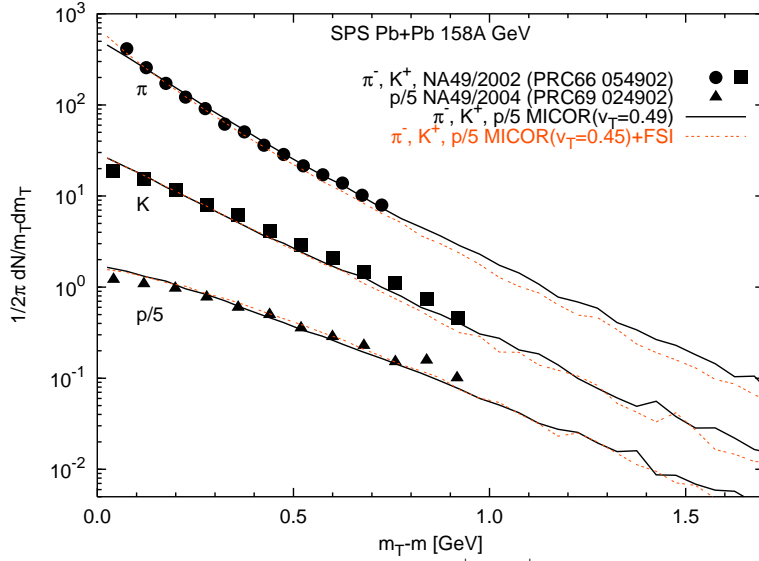}
\vspace*{-8mm}
\caption[]{Transverse momentum spectra for $\pi^+, \ K^+$, and $p$
at $\sqrt{s_{NN}}= 17.3\,\mathrm{GeV}$ in PbPb collisions
(dots, squares and triangulars) \cite{na49-pi+K,na49-p}.
Full lines indicate  quark coalescence results at $v_T=0.60$,
neglecting final state interactions. 
%Thin dotted lines indicate the MICOR results at a smaller transverse flow,
%$v_T=0.55$, but including final state interactions.
}
\label{fig-slopes0-SPS}
\end{figure}

The number of secondary collisions suffered by a particle
moving with velocity $v$ can be determined using the following rough estimate:
\begin{equation}
N_\mathrm{coll}(v)\,\ \sim\,\ \intop_{t_0}^\infty n(t)\sigma\, v\, dt
\,\ \sim\,\ \intop_{t_0}^\infty
{C'\over t^\alpha \bar v_\mathrm{flow}}\,\sigma\,v\,dt
\,\ =\, {C\,\sigma\, v\over\bar v_\mathrm{flow}}\,\ ,
\end{equation}
where $n(t)$ is the density of the matter at the location of the
particle at time $t$ and $\sigma$ denotes the averaged cross section. 
The density, $n(t)$, is supposed to be inversely proportional to the
average flow velocity $\bar v_\mathrm{flow}$ and a power of the elapsed time
$t^\alpha$. If the exponent $\alpha$ is a constant, then $C$ is also
constant. The average flow velocity in our case can be approximated by the
transverse quark flow velocity $v_T$. 
The result of this simple calculation is that a particle 
suffers at least one collision ($N_{coll}(v) \geq 1$)
if its individual velocity $v$ 
satisfies the following relation:
\begin{equation}
v\,\ \geq\,\ v_T^\mathrm{flow}/C\sigma.
\end{equation}

Now switching on final state interactions (thin dotted lines)
and focusing on proton spectra, we are able to reproduce  proton data
with a smaller quark flow superposed with the gained transverse energy
during secondary interactions.
On the other hand, the calculated light meson spectra will deviate from 
data, because the ``pion wind'' decreases the pion yield at higher
momenta. From Figure \ref{fig-slopes-RHIC}
we can claim that there is no free room for 
final state interactions.  Assuming fast hadronization, we are able
to reproduce experimental data. 
However, if we want to include light mesons produced via
independent jet-fragmentation, then the second case with final state
interaction looks more favourable,
where extra meson yields can be accommodated.  
(Contribution from jet fragmentation appears at much higher $p_T$
for protons.) At this point we can not decide between the two
investigated scenarios.
Further analysis of perturbative QCD
contributions is needed, similarly to Refs.~\cite{recHwa,recFries,coalGKL}
and Ref.~\cite{Zhang02}.

\newpage

\begin{figure}[ht]
                 \insertplot{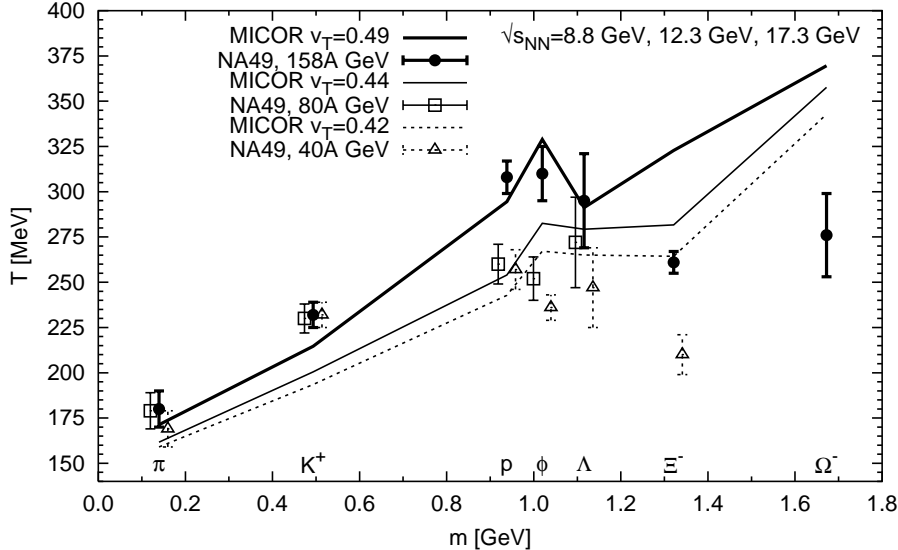}
\vspace*{-8mm}
\caption[]{Inverse slopes from MICOR model without final state
   interactions and experimental data from NA49
  \cite{na49-pi+K,na49-p,na49-phi+Omega,na49-Lambda}.
  MICOR points are connected by dotted lines for $v_T=0.42$ flow velocity,
  thin lines for $v_T=0.44$ and thick lines for $v_T=0.49$.
  Experimental data points 
  at $40\,A\,\mathrm{GeV}$ are denoted by open
  triangles, $80\,A\,\mathrm{GeV}$ data are displayed by squares,
  $158\,A\,\mathrm{GeV}$ data are indicated by filled dots.
%  For $40\,A\,\mathrm{GeV}$ and
%  $80\,A\,\mathrm{GeV}$ data points are shifted by 
%  $\Delta m = \pm0.02\,\mathrm{GeV}$.
}
\label{fig-slopes-SPS0}
\end{figure}

Figure \ref{fig-slopes0-RHIC} 
summarizes our results on slope parameters obtained in MICOR,
when final state interactions are neglected. 
The uncertainties in quark transverse flow are displayed
selecting
$v_T=0.55 \pm 0.02$ at $\sqrt{s_{NN}}=130$ GeV (thin lines) and 
$v_T=0.60 \pm 0.02$ at $\sqrt{s_{NN}}=200$ GeV (thick lines). 
The MICOR results agree with the measured slopes quite well, which 
agreement confirms the application of MICOR model and the assumption of
quark-antiquark matter formation at RHIC.

At SPS we expect slower expansion and hadronization, which  yield
to a less energetic collision dynamics. Thus after hadronization, secondary
interactions may have a larger role to reshape the final 
momentum distributions and modify the inverse slopes.
To validate this expectation, we repeat our MICOR calculation
at SPS energies with and without final state interactions.

Figure \ref{fig-slopes0-SPS} displays the MICOR result for
$\pi^+, \ K^+$, and $p$ transverse momentum spectra
at $\sqrt{s_{NN}}=17.3$ GeV \cite{na49-pi+K,na49-p}
in PbPb collisions at SPS. Without final state interactions (full lines)
one can reproduce the pion and proton data at 
$v_T=0.49$. For kaons we find slight 
deviation. Introducing final state interactions and focusing on the
proton spectra, we obtain $v_T=0.45$. Proton spectra is reproduced
in both cases. Kaon and pion spectra are similar  where 
data are available. However, final state interactions modify 
pion and kaon spectra by decreasing the yield at higher $p_T$.
The obtained suppression in the spectra 
at SPS is much less than at RHIC.

\newpage

\begin{figure}[htb]
                 \insertplot{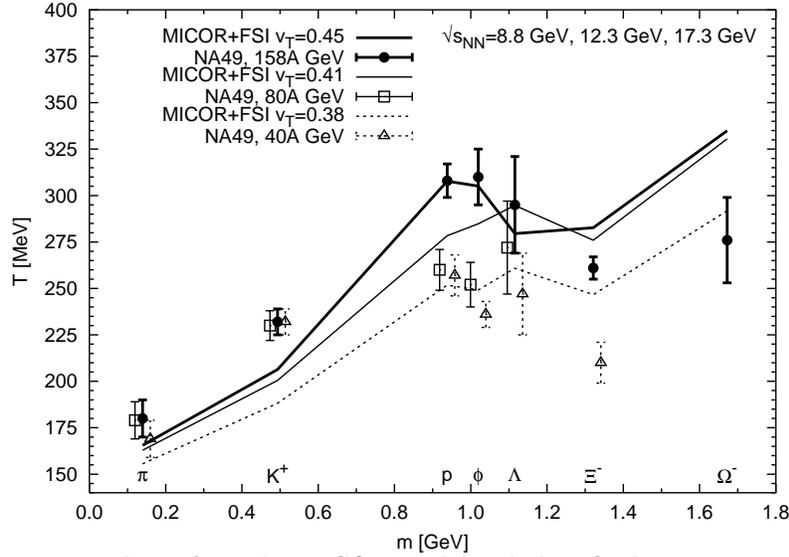}
\vspace*{-8mm}
\caption[]{Inverse slopes from the MICOR model including final state
   interactions on NA49 experimental data
  \cite{na49-pi+K,na49-p,na49-phi+Omega,na49-Lambda}.
  MICOR points are connected by dotted lines for $v_T=0.38$ flow velocity,
  thin lines for $v_T=0.40$ and thick lines for $v_T=0.44$.
  $40\,A\,\mathrm{GeV}$ experimental data points are denoted by open
  triangles, $80\,A\,\mathrm{GeV}$ points are open squares,
  $158\,A\,\mathrm{GeV}$ points are filled circles.
  }
\label{fig-slopes-SPS}
\end{figure}

The displayed differences between data and the MICOR calculation with
final state interaction indicate that hadronic collisions 
can become important in a wide energy range.
However, we need to investigate whether pQCD  processes
(namely independent jet fragmentation)
could provide the missing contributions both at RHIC and SPS energies.

Now we summarize MICOR results on transverse slopes
at projectile energies
$E_{beam}=40, \ 80, \ 158$ AGeV at SPS.
Figure \ref{fig-slopes-SPS0} shows the slopes in case of 
neglecting final state interactions. 
Figure \ref{fig-slopes-SPS} 
displays the slopes for including final state effects.
In the latter case experimental data
can be reproduced by transverse flow $v_T^{(FSI)}=0.38,\  0.40,\  0.44$,
respectively.  These $v_T$ values are
smaller than in the interactionless case, the difference is $v_T^{(no FSI)} -
v_T^{(FSI)} \simeq 0.04$.  It means that final state interactions increase
transverse flow effect. Thus we can start from smaller quark flow
values for the expanding quark matter, after switching on hadronic interactions
the hadronic transverse momentum spectra will be reproduced.

%This picture suggests that for low-$p_T$ particles
%final state interactions are important 
%at small expansion rate, only. This happens
%at smaller nuclear collision energies, namely at SPS energies.
%One can determine the appropriate transverse flow values by 
%the GROMIT model, which follows secondary collisions.

Inverse slopes on Figure ~\ref{fig-slopes-SPS} are fitted at low $p_T$
($p_T < 1$ GeV).
There is a small difference between data and calculations at
smaller collision energies, even for the best choice of the $v_T$ values.
Pion and kaon inverse slopes from MICOR are significantly higher than 
data at $E_{beam} =40, \ 80$ AGeV.
At larger energy, $E_{beam} = 158$ AGeV,
the agreement is recovered using $v_T=0.44\pm0.02$.

\newpage

\begin{figure}[htb]
                 \insertplot{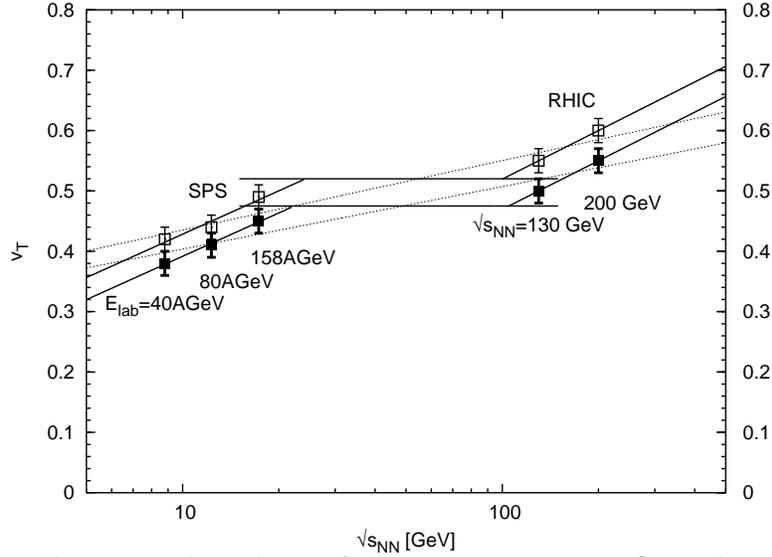}
\vspace*{-8mm}
\caption[]{The energy dependence of the average transverse flow velocities
           of the expanding quark matter in the MICOR model
           including final state effects (full boxes) and 
           neglecting them (open boxes).
  }
\label{fig-vt}
\end{figure}

Now the results of  Figures \ref{fig-slopes0-RHIC}, 
\ref{fig-slopes-SPS0} and 
\ref{fig-slopes-SPS} 
can be combined into one figure  to display 
energy dependence of transverse quark flow, $v_T(\sqrt{s_{NN}})$,
see Figure \ref{fig-vt}.
Including final state interactions (full boxes) we obtain 
smaller values for the quark flow than neglecting them in a fast
hadronization scenario (open boxes).
The uncertainties on the transverse flow values allow us to construct
two interpretations.

First we can assume a simple logarithmic energy dependence (dotted lines).
If we neglect final state effects, then 
we can draw the following line to guide the eyes
\begin{eqnarray}
v_T(\sqrt{s_{NN}}) &=& 0.32 + 0.050 \cdot \ln (\sqrt{s_{NN}}) 
\ \ .
\end{eqnarray}
Including final state effects the energy 
dependence of the quark transverse flow is 
\begin{eqnarray}
v_T(\sqrt{s_{NN}}) &=& 0.30 + 0.045 \cdot \ln (\sqrt{s_{NN}})
\ \ .
\end{eqnarray}
Here the energy $\sqrt{s_{NN}}$
is measured in GeV.

On the other hand, our results obtained in discrete energy points
may give a hint for a plateau between the
largest CERN SPS and the smallest accessible RHIC energies (solid lines). 
Further studies are needed to decide
if this hypothetical plateau exists at all and what are the
theoretical consequences. The existence of such a plateau
could yield important information about the equation of state
of deconfined matter.

The increasing transverse quark flow can not violate causality, thus it will
saturate. However, this saturation seems to appear well beyond RHIC energies.

\newpage
\section{Conclusions}\label{concl}

In this paper we summarized our results on the energy
dependence of quark transverse flow. We assumed that deconfined matter
was produced at SPS and RHIC energies in central Pb+Pb and Au+Au collisions
and an initial quark transverse flow has been developed before hadronization.
Using the MICOR hadronization model we could extract quark transverse flow
parameters at phase transition. The comparison of calculated
hadronic transverse momentum spectra and measured ones yields quark flow
values with high precision in a wide energy region.
Assuming fast hadronization and neglecting final state hadronic
collisions, one can determine the wanted energy dependence for quark flow.
We are pleased to see the good agreement between data and our quark based hadronization
model calculations at low-$p_T$ ($p_T < 1$ GeV).

Using the GROMIT transport code, one can investigate the influence
of final state hadronic effects. In this case smaller quark flow values 
can be used to reproduce experimental data.
At first we reproduce the proton transverse momentum
spectra. Applying the extracted quark flow, the pion and kaon
spectra will be smaller than the experimental ones. 
However, this deviation allows us to
include  meson production from independent jet fragmentation,
determined by perturbative QCD calculations.
These perturbative contributions become comparable to the quark
coalescence yield in the region $ p_T \geq 1 $ GeV for pions and kaons.
A detailed perturbative QCD calculation is needed to  determine
the interplay between perturbative, quark coalescence and hadronic
contributions for light mesons.
For heavy baryons and antibaryons
the perturbative contributions are negligible in the region
$p_T \leq 4$ GeV, where coalescence and secondary hadronic
interactions dominate their momentum spectra.

The energy dependence of the obtained quark transverse
flow values can be described by a simple logarithmic increase.
On the other hand, the high precision of the extracted flow values
may indicate the possible existence of a plateau between SPS and available
RHIC energies. Further analysis of the existing experimental
data is needed to confirm this plateau. In parallel, new
data would be needed in the energy range
$20 \leq \sqrt{s_{NN}} \leq 130$ GeV to determine the energy dependence more
precisely.

\section*{Acknowledgments}
The authors are grateful to Gyuri Wolf for organizing the Budapest'04
Workshop and for providing an excellent scientific and social
atmosphere. We thank Brian A. Cole for his comments.
This work was supported by Hungarian OTKA grant T043455, T038406.

\vfill\eject
\end{document}